# NEGATIVE ENERGIES AND FIELD THEORY

**Gerald E. Marsh**

Argonne National Laboratory (Ret)
5433 East View Park
Chicago, IL 60615

E-mail: gemarsh@uchicago.edu

**Abstract:** The assumption that the vacuum is the minimum energy state, invariant under unitary transformations, is fundamental to quantum field theory. However, the assertion that the conservation of charge implies that the equal time commutator of the charge density and its time derivative vanish for two spatially separated points is inconsistent with the requirement that the vacuum be the lowest energy state. Yet, for quantum field theory to be gauge invariant, this commutator *must* vanish. This essay explores how this conundrum is resolved in quantum electrodynamics.

**Introduction**

The assumption that the vacuum is the minimum energy state, invariant under unitary transformations, is fundamental to quantum field theory. However, Schwinger [1] long ago pointed out that there is a problem: The assertion that the conservation of charge implies that the equal time commutator of the charge density and its time derivative vanish for two spatially separated points is inconsistent with the requirement that the vacuum be the lowest energy state. Such commutators are referred to as Schwinger terms. Solomon [2] has also shown that for QFT to be gauge invariant this Schwinger term must vanish. He does this by considering a simplified field theory of non-interacting fermions acted on by a classical electromagnetic field. The following discussion of the Schwinger terms is also in this context.

The existence of non-zero Schwinger terms also impacts Lorentz invariance. Lev [3] has shown that if the Schwinger terms do not vanish the usual current operator $\hat{J}^\mu(x)$, where $\mu = 0, 1, 2, 3$ and $x$ is a point in Minkowski space, is not Lorentz invariant.

The cognoscenti are aware of much that is laid out below. Nonetheless, because the idea of negative energy states is still quite controversial and much confused in the literature, it is hoped that what follows may prove helpful.

**Schwinger Terms**

To begin with, it is important to understand the details of Schwinger's argument. Using Solomon's notation, the Schwinger term is given by

$$\text{ST}(\vec{y},\vec{x}) = \left[\hat{\rho}(\vec{y}), \hat{\vec{J}}(\vec{x})\right]. \tag{1}$$

Taking the divergence of the Schwinger term and using the relation

$$i\left[\hat{H}_0, \hat{\rho}(\vec{x})\right] = -\nabla \cdot \hat{\vec{J}}(\vec{x}), \tag{2}$$

results in

$$\nabla_{\vec{x}} \cdot \left[\hat{\rho}(\vec{y}), \hat{\vec{J}}(\vec{x})\right] = \left[\hat{\rho}(\vec{y}), \nabla \cdot \hat{\vec{J}}(\vec{x})\right] = -i\left[\hat{\rho}(\vec{y}), \left[\hat{H}_0, \hat{\rho}(\vec{x})\right]\right]. \tag{3}$$

(The argument being given here is not restricted to the free-field Hamiltonian $\hat{H}_0$. For an explanation of how the free-field Hamiltonian makes its appearance in Eq. (2), see [2].)

Expanding the commutator on the right hand side of Eq. (3) yields the vacuum expectation value

$$i\nabla_{\vec{x}} \cdot \left\langle 0 \left| \left[ \hat{\rho}(\vec{y}), \hat{\vec{J}}(\vec{x}) \right] \right| 0 \right\rangle = -\left\langle 0 \left| \hat{H}_0 \, \hat{\rho}(\vec{x})\hat{\rho}(\vec{y}) \right| 0 \right\rangle +$$
$$\left\langle 0 \left| \hat{\rho}(\vec{x})\hat{H}_0\hat{\rho}(\vec{y}) \right| 0 \right\rangle + \left\langle 0 \left| \hat{\rho}(\vec{y})\hat{H}_0\hat{\rho}(\vec{x}) \right| 0 \right\rangle - \left\langle 0 \left| \hat{\rho}(\vec{y})\hat{\rho}(\vec{x})\hat{H}_0 \right| 0 \right\rangle.$$
(4)

It is here that one makes the assumption that the vacuum is the lowest energy state. This done by writing $\hat{H}_0|0\rangle = \langle 0|\hat{H}_0 = 0$. As a result, Eq. (4) may be written as

$$i\nabla_{\vec{x}} \cdot \left\langle 0 \left| \left[ \hat{\rho}(\vec{y}), \hat{\vec{J}}(\vec{x}) \right] \right| 0 \right\rangle = \left\langle 0 \left| \hat{\rho}(\vec{x})\hat{H}_0\hat{\rho}(\vec{y}) \right| 0 \right\rangle +$$
$$\left\langle 0 \left| \hat{\rho}(\vec{y})\hat{H}_0\hat{\rho}(\vec{x}) \right| 0 \right\rangle.$$
(5)

Multiply both sides of the last equation by $f(x)f(y)$ and integrate over $x$ and $y$. The right hand side of Eq. (5) becomes

$$\int d\vec{x}\, d\vec{y} \left\{ \left\langle 0 \left| f(\vec{x})\hat{\rho}(\vec{x})\hat{H}_0 f(\vec{y})\hat{\rho}(\vec{y}) \right| 0 \right\rangle + \left\langle 0 \left| f(\vec{y})\hat{\rho}(\vec{y})\hat{H}_0 f(\vec{x})\hat{\rho}(\vec{x}) \right| 0 \right\rangle \right\}.$$
(6)

If Schwinger's "arbitrary linear functional of the charge density" is defined as

$$F = \int f(\vec{x})\hat{\rho}(\vec{x})d\vec{x} = \int f(\vec{y})\hat{\rho}(\vec{y})d\vec{y},$$
(7)

the right hand side of Eq. (5) becomes

$$2\left\langle 0 \left| F\hat{H}_0 F \right| 0 \right\rangle = 2\sum_{m,n} \left\langle 0 \left| F \right| m \right\rangle \left\langle m \left| \hat{H}_0 \right| n \right\rangle \left\langle n \left| F \right| 0 \right\rangle =$$
$$2\sum_n E_n \left\langle 0 \left| F \right| n \right\rangle \left\langle n \left| F \right| 0 \right\rangle = 2\sum_n E_n \left| \left\langle 0 \left| F \right| n \right\rangle \right|^2 > 0.$$
(8)

The left hand side of Eq. (8)—essentially the form used by Schwinger—is here expanded to explicitly show the non-vanishing matrix elements between the vacuum and the other states of necessarily positive energy. *This shows that if the vacuum is assumed to be the lowest energy state, the Schwinger term cannot vanish, and the theory is not gauge invariant.* The above argument is based on that given by Solomon, who also shows the converse, that if the Schwinger term vanishes, then the vacuum is not the lowest energy state and the theory *is* gauge invariant.

For the sake of completeness, it is readily shown that the left side of Eq. (5) becomes

$$i\int \nabla_{\vec{x}} \cdot \left\langle 0 \left| \left[ \hat{\rho}(\vec{y}), \hat{\vec{J}}(\vec{x}) \right] \right| 0 \right\rangle f(\vec{x})f(\vec{y}) \mathrm{d}\vec{x}\mathrm{d}\vec{y} = i \left\langle 0 \left| [\partial_t F, F] \right| 0 \right\rangle, \tag{9}$$

so that combining Eqs. (8) and (9) yields a somewhat more explicit form of the result given by Schwinger,

$$i \left\langle 0 \left| [\partial_t F, F] \right| 0 \right\rangle = 2\sum_n E_n |\langle 0| F |n\rangle|^2 > 0. \tag{10}$$

One way to resolve the difficulty raised by Schwinger terms is to define the product of localized field operators as the singular limit of products defined at separate points; i.e., to introduce "point splitting". While this is often done, point splitting has its own problems having to do with the definition of the current operator, the equations of motion for the fields, and the Lorentz invariance of the theory. These issues have not yet been resolved, as discussed by Boulware [4] and Solomon [5].

The issue of the gauge invariance of QFT, raised by Solomon, has been dealt with in a variety of ways over the years (an extensive discussion is contained in [2]). In essence, the standard vacuum of QFT is only gauge invariant if non-gauge invariant terms are removed. There are two general approaches to achieving this: the first is to simply ignore such terms as being physically untenable and remove them so as to maintain gauge invariance; and the second is to use various regularization techniques to cancel the terms.

In QFT, relativistic transformations between states are governed by the continuous unitary representations of the inhomogeneous group SL(2,C)—essentially the complex Poincaré group. One might anticipate that when interactions are present the unitarity condition might be violated. Indeed, Haag's theorem [6] states, in essence, that if $\phi_0$ and $\phi$ are field operators defined respectively in Hilbert spaces $\mathcal{H}_0$ and $\mathcal{H}$, with vacua $|0\rangle_0$ and $|0\rangle$, and if $\phi_0$ is a free field of mass $m$, then a unitary transformation between $\phi_0$ and $\phi$ exists only if $\phi$ is also a free field of mass $m$. Another way of putting this is that if the interaction picture is well defined, it necessarily describes a free field.

Today it is well known that the physical vacuum state is not simple and must allow for spontaneous symmetry breaking and a host of other properties, so that the real vacuum bears little relation to the vacuum state of axiomatic QFT. Nevertheless, even if the latter type of vacuum is assumed, the violation of the unitarity condition in the presence of interactions opens up the possibility that the spectral condition, which limits momenta to being within or on the forward light cone, may also be violated thereby allowing negative energy states. As will be discussed below, in quantum electrodynamics (QED) transitions to negative energy states are explicitly allowed.

Of course, the way QFT gets around the formal weakness of using the interaction picture is to regularize the singular field functions that appear in the perturbation series followed by renormalization. There is nothing wrong with this approach from a pragmatic point of view, and it works exceptionally well in practice.

**Feynman Diagrams**

In what follows, Feynman diagrams will be used to explore the issue of negative energy states in QED. Before entering into this discussion, however, it is important to explore the implicit assumptions made when using these diagrams.

Feynman [7] in his famous paper "The Theory of Positrons" observed that the Schrödinger and Dirac equations can be visualized as describing the scattering of a plane wave by a potential. The scattered waves may proceed both forward and backward in time and may suffer further scattering by the same or other potentials. An identity is made between the negative energy components of the scattered wave and the waves traveling backward in time. This interpretation is valid for both virtual particles (that appear in the use of Feynman diagrams to give a graphical representation of a perturbation series) and for real particles, where the energy-momentum relation $E^2 = p^2c^2 + m^2c^4$ must be satisfied. While one generally does not indicate the waves, and instead draws world-lines in Minkowski space between such scatterings, it is generally understood that the particle represented by these waves does not have a well defined location in space or time between scatterings [8].

The Feynman approach visualizes a non-localized plane wave impinging on a region of spacetime containing a potential, and the particle the wave represents being localized [9] to a finite region of Minkowski space by interaction with the potential. The waves representing the scattered particle subsequently spread through space and time until there is another interaction in the same potential region or in a different region also containing a potential, again localizing the particle. Even this picture is problematic since the waves are not observable between interactions. The figure below is intended to represent electron scattering from two different locations in a region containing a scattering potential. A plane electron wave (1) comes in from the lower left of the figure, is scattered by the potential at A(3). (a) shows that the scattered wave can go both forward and backward in time; (b) and (c) show two second order processes where (b) shows a normal scattering forward in time and (c) the possibility of pair production. Feynman meant this figure to apply to a virtual process, but—as mentioned earlier—can apply as well to real pair production provided the energy-momentum relation is satisfied. Although lines are drawn to represent the paths of these particles, no well-defined world-lines exist.

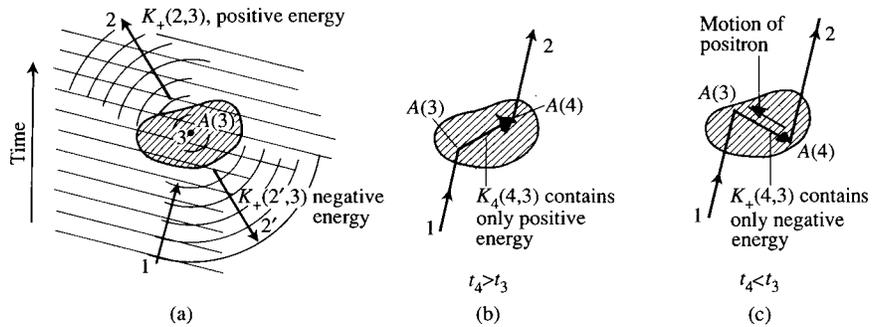

Figure 1. Propagation kernels associated with the Dirac equation. [Based on Figure 2 of R. P. Feynman, "Theory of Positrons", *Phys. Rev.* **76**, 749-759 (1949)]

In a particle detector such as a bubble chamber, where the paths followed by the charged particles are made visible by repeated localizing interactions with a medium, one would observe in (c) a pair creation event at A(4), an electron coming in from the lower left, and an annihilation event at A(3). Of course, since the particles involved here are massive, in the case of real pair production the interval between A(3) and A(4) is time-like and the spatial distance between these events depends on the observer. If the interval between these points is space-like, and this will be relevant below, the time ordering of A(3) and A(4) depends on the observer.

To reiterate, a world-line is a classical concept that is only approximated in quantum mechanics by the kind of repeated interactions that make a path visible in a particle detector such as a bubble chamber [10]. Minkowski space is the space of *events*—drawing a world-line in a Minkowski diagram implicitly assumes such repeated interactions taken to the limit of the continuum [11]. While the characterization of Minkowski space as the space of events is often obscured by drawing world-lines as representing the putative path of a particle in spacetime independent of its interactions, remembering that—in the context of quantum mechanics—each point in Minkowski space is the position of a potential event removes much of the apparent incompatibility between quantum mechanics and special relativity.

The picture one now has, however, is very unlike that of the path of a massive particle—like a marble—moving in spacetime. Consider a Minkowski diagram showing the world-lines of several marbles at different locations. Given a spacelike hypersurface corresponding to an instant of time in some frame, all the marbles would be visible at some set of locations. If one chooses a neighboring instant of time, these marbles would all still be visible at slightly different locations. This is because of the sharp localization of the marbles in space and time due to the continual interactions of their constituent components. Now consider the case of several elementary particles such as electrons. On any spacelike hypersurface, the only particles "visible" would be those that were localized by an interaction to a region of spacetime that included the instant of time corresponding to the hypersurface [12]. After any localization, the wave function of a particle spreads both in space and in either direction in time. Consequently, neighboring hypersurfaces (in the same reference frame) corresponding to slightly different times could have a different set of particles that were "visible." If motion consists of a sequential series of localizations along a particle's path, it is not possible to define a continuum of movement in the classical sense—there exists only a series of "snapshots."

Haag, [13] has put this somewhat different terms: "The resulting ontological picture differs drastically from a classical one. It sketches a world, which is continuously evolving, where new facts are permanently emerging. Facts of the past determine only probabilities of future possibilities. While an individual event is considered as a real fact, the correlations between events due to quantum mechanical entanglement imply that an individual object can be regarded as real only insofar as it carries a causal link between two events. The object remains an element of potentiality as long as the target result has not become a completed fact."

It is important to emphasize that between localizations due to interactions, an elementary particle does not have a specifiable location, although—because it has a high probability of being located somewhere within the future and past light cones associated with its most recent localization—it would contribute to the local mass-energy density. The lack of a definite location is not a matter of our ignorance, it is a fundamental property of quantum mechanics; Bell's theorem tells us that there are no hidden variables that could specify a particle's position between localizations.

As an example of how localization works, consider a single atom. Its nucleus is localized by the continuous interactions of its constituent components. The electrons are localized due to interactions with the nucleus, but only up to the appropriate quantum numbers—*n, l, m*, and *s*. One cannot localize the electrons to positions in their "orbits."

**Quantum Electrodynamics, Negative Energy, and Charge Conservation**

In QED, the kernel that propagates an electron forward in time uses only positive energies. As a consequence, *the amplitude cannot vanish anywhere outside the light cone* [14], although it becomes small over a distance comparable to the Compton wavelength. As a result, not much of the space outside the light cone is really accessible. This is the basis of Feynman's well known Dirac lecture on the reason for antiparticles [15].

The figure shown below is meant to apply to real processes rather than virtual ones so that both energy and momentum must be conserved. This means that external potentials (often explicitly represented by a wavy line with a cross at the end) must be present at A(3) and A(4) for the indicated interactions to occur since they are otherwise kinematically forbidden.

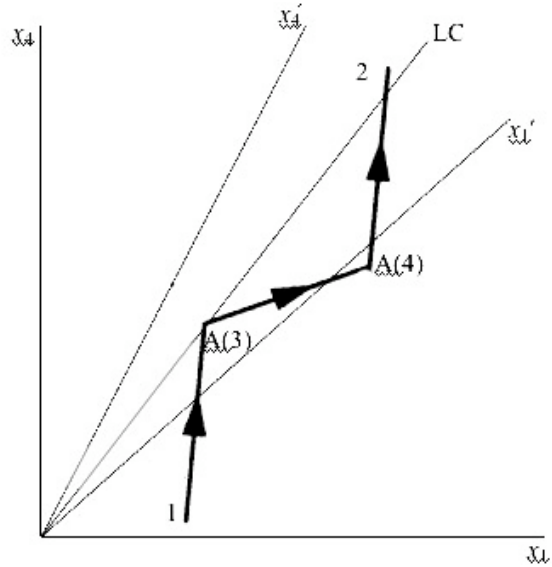

Figure 2. Kernel that propagates an electron forward in time uses only positive energies and consequently cannot vanish outside the light cone (LC). As indicated, the electron is scattered by the external potentials at A(3) and A(4).

In the Lorentz frame labeled ($x_1$, $x_4$), the diagram may be interpreted as depicting an electron being propagated from 1 to A(3) where it is scattered by interaction with an external potential and again scattered at A(4) after which it is propagated to 2. The "path" from A(3) to A(4) is taken to be outside the light cone of the scattering event at A(3). This is possible since the propagation kernel from 1 to A(3) contains only positive energies and therefore—as was mentioned earlier—cannot vanish outside the light cone.

In the Lorentz frame ($x_1'$, $x_4'$) there are two ways to interpret the diagram. The first is where an electron is propagated from 1 to A(3) and pair creation occurs at A(4). The positron is then propagated from to A(4) to A(3) where it annihilates the electron propagated from 1 to A(3) with the emission of a photon, and the electron from the pair creation event at A(4) is propagated to 2.

Alternatively, the electron at 1 is propagated to A(3) where it transitions to a negative energy state as a result of interacting with the external potential and is subsequently propagated backwards in time to A(4). In Feynman's interpretation, a particle in a negative energy state (here an electron) propagating backwards in time is equivalent to an anti-particle having positive energy (here a positron) propagating forwards in time. At A(4) the electron moving backwards in time interacts with the external potential so as to raise it to a positive energy state and is then propagated to 2. Note that on a space-like hypersurface such as $x_1'$, charge is conserved because an electron moving backwards in time has a positive charge. Another way of saying this, following Weinberg [16], is that—in the context of quantum field theory—if an electron in an external field obeys the quantized [17] Dirac equation, one cannot rule out the negative energy solutions needed to make up a complete set of wave functions. But a wave function representing an electron in a negative energy state can only be non-zero if it has charge $+e$.

There is an easy way to see that this using only relativistic quantum mechanics. If $\Psi(x)$ represents the motion of a Dirac particle of mass $m$ and charge $e$ in a potential $A_\mu(x)$, then it satisfies the equation

$$[\gamma^\mu(i\partial_\mu - eA_\mu) - m]\Psi(x) = 0.$$

(11)

On the other hand $\Psi^C(x) = K_C\Psi(x)$, where $K_C$ is the charge conjugation operator [18], satisfies the Dirac equation for a particle of the same mass but opposite charge in the same potential. Here $K_C \equiv \gamma^5 K$ and $K$ is an anti-unitary operator such that $K^2 = -1$. Both the operator $r$ and the Dirac matrices $\gamma^\mu$ are conserved under $K$ while the operator $p$ is transformed to $-p$. Most importantly, if the average value of the Hamiltonian in the state $\Psi$ is $\langle \mathcal{H} \rangle = \langle \Psi | \mathcal{H} | \Psi \rangle$ then one can show that this average value is related to that of the charge conjugate state $\Psi^C$ by

$$\langle \mathcal{H}(-e) \rangle_C = -\langle \Psi | \mathcal{H}(e) | \Psi \rangle,$$

(12)

so that charge conjugation also changes the sign of the energy. Two charge-conjugate solutions have the same probability and current densities and therefore opposite charge and electric-current densities.

Look again at Fig. 1(c). This figure is essentially the same as what is seen in the Lorentz frame $(x_1', x_4')$ of Fig. 2. Thus, the appropriate propagation kernel between A(3) and A(4) in Fig. 2 is $K_+(4, 3)$, which contains only negative energy. This is the case despite the fact that we started with a propagation kernel from 1 to A(3) that contained only positive energy. This is a consequence of the "path" from A(3) to A(4) being outside the light cone of the scattering event at A(3) in the $(x_1, x_4)$ frame. The same is true for Fig. 1(c) since any scattering backwards in time must be outside the light cone of the scattering event.

Figure 1(c) can be somewhat misleading in that while $K_+(4, 3)$ contains only negative energy, the real energy is actually positive since in the phase $\exp[-iE_n(t_4 - t_3)]$ associated with $K_+(4, 3)$, the energy $E_n$ is indeed negative, but so is $(t_4 - t_3)$ so that the real energy is still positive—given the usual convention where positive energy states evolve in time as $e^{-i\omega t}$.

What remains counterintuitive here is that for propagation outside the light cone, whether we see an electron or a positron depends on our frame of reference. The charge of the electron when it is between A(3) and A(4) can vary in different Lorentz frames: if it is negative in the $(x_1, x_4)$ frame, it is positive in the $(x_1', x_4')$ frame. This challenges the classical idea that charge is an intrinsic and invariant property of a particle. Note again, that on a space-like hypersurface such as $x_1'$, charge is conserved *only* if the electron moving backward in time reverses its charge. Thus, as argued by Feynman, if we insist that propagation kernels use only positive energies, anti-particles must exist if charge is to be conserved.

**Summary**

We have seen that quantum field theories that require that the vacuum state |0> be the state of minimum energy—so that negative energies are not allowed, are plagued by Schwinger terms that destroy gauge invariance. This leads to the panoply of techniques used to eliminate the gauge-invariance destroying terms, essentially some form of regularization or simply ignoring the terms as being unphysical.

On the other hand, quantum electrodynamics, which explicitly allows and makes use of negative energy states, *is* gauge invariant. The Feynman space-time approach, where negative-energy particle solutions propagating into the past are equivalent to positive-energy anti-particle solutions propagating forward in time, gives the same calculative results as Dirac hole theory [7]. In the space-time approach, pair annihilation can be viewed as an electron interacting with a potential and making a transition to a negative energy state by emitting a photon and subsequently propagating backward in time—thereby being equivalent to a positron moving forward in time.

The Feynman approach does not make use of the Dirac idea that the negative energy states are filled with particles obeying the Pauli exclusion principle. As a result, the Feynman approach is also conceptually applicable to particles having spin zero. What one finds in general is that the emission (absorption) of an anti-particle having 4-momentum $p^\mu$ is equivalent to the absorption (emission) of a particle of 4-momentum $-p^\mu$.

Restricting propagation kernels to only positive energies means that anti-particles—which are particles in negative energy states moving backwards in time—must exist if charge is to be conserved.

one. Each of these packets can then serve as a possible starting point for a new trajectory, but each of these starting points must be considered as a separate and distinct possibility, which, if realized, excludes all others." If the particle has large momentum, ". . . the uncertainty in momentum introduced as a result of the interaction with the atom results in only a small deflection, so that the *noninterfering* packets all travel with almost the same speed and direction as that of the incident particle." [emphasis in the original]

[11] There is a considerable—and quite interesting—literature dealing with repeated "measurements" of a particle and what is known as "Turing's Paradox" or the "Quantum Zeno Effect." See, for example: B. Misra and E. C. G. Sudarshan, *J. Math. Phys.* **18**, 756 (1977); D. Home and M. A. B. Whitaker, *Ann. of Phys.* **258**, 237 (1997), lanl.arXiv.org, quant-ph/0401164.

[12] The term "visible" is put in quotes as a shorthand for the physical processes involved: the interaction of the particle needed to localize it on the space-like hypersurface and the detection of that interaction by the observer. It should also be emphasized that localization is in both space and *time*. Just as localization in space to dimensions comparable to the Compton wavelength corresponds to an uncertainty in momentum of $\sim mc$, localization in time must be $\geq h/mc^2$ if the uncertainty in energy is to be less than or equal to the rest mass energy. For electrons this corresponds to $\geq 10^{-20}$ second.

[13] Rudolph Haag, *Quantum Theory and the Division of the World, Mind and Matter* **2**, 53 (2004).

[14] For a clear and extensive discussion, see: J. Hilgevoord, *Dispersion Relations and Causal Description* (North-Holland, Amsterdam 1960).

[15] *Elementary Particles and the Laws of Physics: The 1986 Dirac Memorial Lectures*, R. P. Feynman and S. Weinberg, Lecture notes compiled by R. MacKenzie and P. Doust (Cambridge University Press, Cambridge 1987).

[16] S. Weinberg, *The Quantum Theory of Fields* (Cambridge University Press, New York 1995), Vol. 1, Ch. 14.

[17] In quantum field theory, the Dirac wave function is reinterpreted as a time and position dependent operator. When expanded in terms of the solutions to the Dirac equation, the coefficients of the expansion are operators. The positive energy coefficients are operators that destroy electrons, and the negative energy coefficients are operators that create positrons having positive energy. Such a reinterpretation is also applicable to bosons—unlike the Dirac hole theory.

[18] The conventions used are those of A. Messiah, *Quantum Mechanics* (John Wiley & Sons, New York 1962), Vol. II.